\documentclass[]{emulateapj}
\usepackage{apjfonts}
\usepackage{times}
\slugcomment{Version: \today}
\usepackage{url}
\usepackage{color}

\begin{document}
\title{Upper Bound on the First Star Formation History}

\author{Yoshiyuki Inoue\altaffilmark{1}, Yasuyuki T. Tanaka\altaffilmark{2}, Grzegorz M. Madejski\altaffilmark{1} \& Alberto Dom\'inguez\altaffilmark{3}} 

\affil{$^1$Kavli Institute for Particle Astrophysics and Cosmology, Department of Physics, Stanford University and SLAC National Accelerator Laboratory, 2575 Sand Hill Road, Menlo Park, CA 94025, USA}
\affil{$^2$ Hiroshima Astrophysical Science Center, Hiroshima University, 1-3-1 Kagamiyama, Higashi-Hiroshima, Hiroshima 739-8526,
Japan}
\affil{$^3$Department of Physics \& Astronomy, University of California, Riverside, CA 92521, USA}
\email{E-mail: yinoue@slac.stanford.edu}

%%%%%%%%%%
%%    Abstract    %%
%%%%%%%%%%
\begin{abstract}
Our understanding of the nature of the extragalactic background light (EBL) has improved with the recent development of gamma-ray observation techniques. An open subject in the context of the EBL is the reionization epoch, which is an important probe of the formation history of first stars, the so-called Population III (Pop~III) stars. Although the mechanisms for the formation of Pop~III stars are rather well understood on theoretical grounds, their formation history is still veiled in mystery because of their faintness. To shed light into this matter, we study jointly the gamma-ray opacity of distant objects and the reionization constraints from studies of intergalactic gas. By combining these studies, we obtain a sensitive upper bound on the Pop~III star formation rate density as $\dot\rho_{*}(z)<0.01[(1+z)/{(1+7.0)}]^{3.4}({f_{\rm esc}}/{0.2})^{-1}({C}/{3.0})\ {\rm M}_{\odot}\ {\rm yr}^{-1}\ {\rm Mpc}^{-3}$ at $z\ge7$, where $f_{\rm esc}$ and $C$ are the escape fraction of ionizing photons from galaxies and the clumping factor of the intergalactic hydrogen gas. This limit is a $\sim10$ times tighter constraint compared with previous studies that take into account gamma-ray opacity constraints only. Even if we do not include the current gamma-ray constraints, the results do not change. This is because the detected gamma-ray sources are still at $z\le4.35$ where the reionization has already finished.
\end{abstract}


%%%%%%%%%%%%
%%    Introduction    %%
%%%%%%%%%%%%
\section{Introduction}
\label{sec:intro}
First stars, the so-called Population III (Pop~III) stars, are believed to have played an important role in the early cosmic evolution by emitting intense radiation and by dispersing heavy elements into the interstellar medium \citep[e.g.][]{bro09}. The formation process of Pop~III stars has been theoretically studied in detail. Pop~III stars were formed when the Universe was less than a few hundred million years old \citep[e.g.][]{abe02} and their typical mass was $\sim20-40\  M_\sun$ \citep{hos11,hos12}\footnote{Pop~III stars are divided into Pop~III.1 and III.2 \citep{bro09}. Pop~III.1 stars are stars formed from cosmological initial conditions, and Pop~III.2 stars are zero-metallicity stars but formed from a gas influenced by earlier stars. For simplicity, we treat them as one population.}. However, the star formation history of Pop~III stars is highly uncertain. The theoretically predicted peak of the star formation rate of Pop~III stars varies by four orders of magnitude, over $\sim10^{-5} -10^{-1}\ M_\sun \ \rm{ yr^{-1} \ \rm{Mpc^{-3}}}$ \citep[e.g.][]{bro06, tor07, joh08, tre09, des11}. Observationally, it is difficult to directly investigate Pop~III stars with current telescopes because Pop~III stars are formed at the early stage of the universe's history. Direct measurements will require next generation instruments, as sensitive as the thirty meter telescope\footnote{\url{http://www.tmt.org}} and the James Webb Space Telescope\footnote{\url{http://www.jwst.nasa.gov}}.

There are various indirect lines of evidence for Pop~III stars -- such as reionization \citep[e.g.][]{tot06,ade13}. Intense ultraviolet (UV) radiation from Pop~III stars was very likely to significantly contribute to the ionization of the intergalactic medium (IGM). Measurements of IGM absorption signatures in the spectra of distant quasars and gamma-ray bursts (GRBs), together with the polarization of the cosmic microwave background (CMB), already prove that the IGM has been reionized between redshift $z\sim6$ and $\sim30$ \citep[e.g.][]{mes10,pri10,ade13}. 

The Infrared Telescope in Space ({\it IRTS}) reported an excess in the NIR background, which can be explained by redshifted light from Pop~III stars \citep[][see also \citet{tsu13}]{mat05}. However, the required ionizing photon budget to explain this NIR excess would overproduce the measured Thomson scattering optical depth \citep{mad05}. Later it was also found that this excess would be inconsistent with TeV observations of nearby blazars (e.g. Aharonian~et.~al.~2006, but see also Essey \& Kusenko~2010). 

Gamma rays propagating through the universe are attenuated by pair production interactions with low-energy photons of the extragalactic background light \citep[EBL; e.g.][]{gou66,jel66,dwe13}. We can probe the EBL by measuring the attenuation features in the spectra of distant gamma-ray sources \citep[e.g.][]{aha06, ack12_ebl,abr13_ebl,dom13}. UV radiation fields causing the reionization may induce significant gamma-ray absorption above a few tens of GeV \citep{oh01,sin10,ino13}.

Based on a galaxy formation model that reproduces the reionization history, it has been shown that it is difficult to distinguish the contribution of Pop~III stars through gamma-ray absorption \citep[e.g.][]{ino13}. On the other hand, without addressing the implications for reionization, Pop~III star formation rate density has been constrained  through gamma-ray absorption \citep{rau09,gil11} or the NIR background \citep{fer06}. \citet{shu08} have investigated the allowed range of Pop~III star formation efficiency using the constraints from the Thomson scattering optical depth, although it depends on the optical depth of IGM after the reionization.

In this Letter, we present new constraints on the Pop~III star formation history, jointly taking account of (1): the Thomson scattering optical depth, (2): neutral fraction of IGM, and (3): observed gamma-ray opacity. Since the NIR background measurement is hampered by uncertain intensity of the zodiacal light, we do not include those constraints. Throughout this Letter, we adopt the standard cosmological parameters of $(h, \Omega_M, \Omega_\Lambda) = (0.7, 0.3, 0.7)$.

%%%%%%%%%%%%
%%    Model    %%
%%%%%%%%%%%%
\section{Constraints on First Star Formation History}
\label{sec:const}
Once the star formation history and spectrum of Pop~III stars are given, we can derive the Thomson scattering optical depth, neutral fraction, and gamma-ray opacity \citep[see][and references therein]{ino13}. Theoretically, the Pop~III star formation history $\dot\rho_{*}(z)$ has been extensively studied \citep[e.g.][]{bro06, tor07, joh08, tre09, des11}. Although all models expect the peak of the Pop~III star formation history at $z\gtrsim6$, the expected history is model dependent. In this Letter, we simply assume the broken power-law shape for the Pop~III star formation history at redshift $z$ as
\begin{equation}
\dot\rho_{*}(z)=\dot\rho_{0}\left[\left(\frac{1+z}{1+z_{\rm peak}}\right)^{\alpha}+\left(\frac{1+z}{1+z_{\rm peak}}\right)^{\beta}\right]^{-1},
\end{equation} 
where $\dot\rho_{0}$ gives the normalization in the units of ${\rm M}_{\odot}\ {\rm yr}^{-1}\ {\rm Mpc}^{-3}$ and $z_{\rm peak}$ is the redshift where the star formation rate of Pop~III stars is maximum. We investigate the allowed parameter space of $(\dot\rho_{0}, z_{\rm peak}, \alpha, \beta)$ by comparing against currently available constraints. Since the reionization has occurred at $z\gtrsim6$, we set $6\le z_{\rm peak}\le20$.  Even if we set $z_{\rm peak}\gtrsim20$, the limit on the Pop~III star formation history does not change significantly. We also set $3\le\alpha\le8$ and $-5\le\beta\le0$ expecting the presence of a peak in the star formation history. If we set a delta function-like star formation history, we can expect higher star formation rate density.

We follow \citet{ino13} for the calculation of the Thomson scattering optical depth, neutral fraction, and gamma-ray opacity. In this Letter, we summarize the key assumptions adopted by us. The cosmic stellar emissivity at a given frequency $\nu$ and redshift $z$ is required to evaluate those values and is given by 
\begin{eqnarray}
\nonumber
j(\nu,z)&=&\int_z^{\infty} \left|\frac{dt}{dz'}\right|dz'f_{\rm esc}\dot\rho_{*}(z') \varepsilon[(1+z')\nu/(1+z), z', z]\\
&&\times\exp[-\tau_{\rm IGM}(\nu,z',z)],
\end{eqnarray}
where $\varepsilon(\nu,z',z)$ is the intrinsic emissivity at frequency $\nu$ at $z$  from stars born at $z'$ in units of erg s$^{-1}$ Hz$^{-1}$ M$_{\odot}^{-1}$ given by stellar population synthesis models and $\tau_{\rm IGM}$ is the attenuation opacity in the IGM, respectively. We adopt the IGM opacity of \citet{yos94}. $f_\mathrm{esc}$ is the escape fraction of photons from galaxies with energy above the threshold for ionization of hydrogen. We use stellar population synthesis models that provide the spectral energy distributions with zero metallicity, namely the model of \citet{sch03}. We adopt the Salpeter initial mass function \citep{sal55} for the mass range of 1-100 $M_\sun$. Recent radiation-hydrodynamics simulations suggest that the typical mass of Pop~III stars is limited to $\lesssim20-40 \ M_\sun$ due to the radiation feedback effects \citep{hos11,hos12}. If there are Pop~III stars with different initial mass  functions, our limit on the Pop~III star formation history will change by a factor equal to the ratio of the integrals of the mass functions. We neglect the interstellar dust extinction effect.  

A key parameter is $f_{\rm esc}$, which we assume here to be 0.2 at all redshifts \citep{yaj11}, unless we note otherwise. Observationally, $f_{\rm esc}\simeq0.05$--0.3 is reported at $z\sim3$ \citep{nes13}, but values at $z\ge4$ have not been measured yet. \citet{ono10} have set upper limits of $f_{\rm esc}\lesssim0.6$ at $z=5.7$ and $f_{\rm esc}\lesssim0.9$ at $z=6.6$ for LAEs. \citet{gon13} independently constrained $f_{\rm esc}$ as $\sim0.5$ at $z=3$ and $\sim0.9$ at $z=6$ using constraints on gamma-ray opacity. Another key parameter to evaluate the reionization is the clumping factor $C=\langle n_H^2\rangle/\bar n_H^2$, where $n_{\rm H}$ is the hydrogen density \citep[see][for details]{ino13}.  At $z\gtrsim6$, $C$ is constrained as $\lesssim3$ by combining hydrodynamical simulations and the measurements of the Ly$\alpha$ forests \citep{bol07}. We take $C=3.0$ as the fiducial value of the parameter unless otherwise noted.

\subsection{Constraint from the Thomson scattering optical depth}
During the reionization epoch, a large population of free electrons is generated in the IGM.  Those electrons scatter the CMB photons, and their optical depth can be estimated from the CMB measurements.  The measured electron Thomson scattering optical depth $\tau_e$ is $\tau_e=0.092\pm0.013$ inferred by the {\it Planck} data \citep{ade13}. Since Population I and II stars also contribute to the reionization, our model must not overproduce the free electron content solely with Pop~III stars.

\subsection{Constraint from the neutral fraction of hydrogen}
As the universe was being reionized, the neutral fraction of intergalactic hydrogen $x_{\rm HI}$ also changed. $x_{\rm HI}$ is a good tracer of the ionizing photon production history during the reionization epoch. The discovery of broad troughs of Ly$\alpha$ absorption, the so-called Gunn-Peterson troughs \citep{gun65}, in the spectra of distant objects have enabled an estimate of the neutral fraction as $x_{\rm HI}\ge0.033$ at $z=6.2$ \citep{mes07}, $x_{\rm HI}\ge0.1$ at $z=7.08$ \citep{mor11}, and $x_{\rm HI}\le0.17$ at $z=6.3$ \citep{tot06}. 

More ionizing photons and lower neutral fraction are expected, if we take into account the Population I and II stars. Therefore, we constrain the Pop~III star formation history not to be less than the lower limit of the neutral fraction. We have not attempted a detailed comparison with Gunn-Peterson measurements of quasars at $z\lesssim6$. Such effects depend rather sensitively on the dense region of the IGM gas, which is not essential for  studying the Pop~III star formation at $z\gtrsim6$.

\subsection{Constraint from the gamma-ray attenuation}
Sufficient UV radiation fields causing cosmic reionization may induce gamma-ray absorption above a few tens of GeV \citep{oh01,sin10,ino13}. To constrain the gamma-ray opacity at high redshifts, we use the {\it Fermi}-LAT detected distant sources with detected $>10$ GeV photons. Since gamma-ray opacity for photons with $\sim20$--70 GeV is close to unity at $z\gtrsim2$ \citep[e.g.][]{dom11,ino13}, we focus on sources at $z\ge2$.

Known distant gamma-ray emitting populations are GRBs and blazars. LAT has detected 35 GRBs \citep{ack13_grb}. Since more distant sources can put tighter constraints on the Pop~III star formation rate, we use the most distant one GRB~080916C at $z=4.35$, which had photons up to 13.22~GeV. We select blazars having Test Statistics (TS) $>100$, photon index $\Gamma\le2.0$, and $z>2.0$ from the second catalog of {\it Fermi}-LAT sources \citep[2FGL][]{nol12_catalog}. After the selection, flat spectrum radio quasar 2FGL~J1344-1723 at $z=2.506$ survives with TS=967.

For both the GRB and blazar cases, we reduced {\it Fermi}-LAT data and performed an unbinned maximum likelihood analysis using Fermi Science Tool version v9r27p1. For 2FGL~J1344-1723, the analyzed time interval is MET~239557417 (2008-08-04 15:43:36 UT) to 386632050 (2013-04-02 21:47:27 UT) representing a total of $\sim$4.7 years of data.  For GRB~080916C, we used the first 1000 seconds after the burst \footnote{\url{http://fermi.gsfc.nasa.gov/ssc/observations/types/grbs/grb\_table/}} (MET~243216766--243217766).  We extracted 0.1--300 GeV {\tt SOURCE} class events within $10^{\rm o}$ circles centered at the source positions. Good time intervals were generated using the recommended filter of {\tt (DATA\_QUAL==1)\&\&(LAT\_CONFIG==1)\&\&ABS(ROCK\\ \_ANGLE)$<$52} and an ROI-based zenith angle cut ({\tt roicut=yes}) with a maximum zenith angle cut of $100^{\rm o}$. We utilized {\tt P7SOURCE\_V6} as the Instrument Response Functions. In the source model of 2FGL~J1344-1723, we modeled the spectrum as a log-parabola and included all the 2FGL sources within a $10^{\rm o}$ circle. The spectral parameters of these sources were left free in the likelihood fitting. 2FGL sources within an annulus of $10^{\rm o}$ to $15^{\rm o}$ from the source were also included in the model, but their spectral parameters were fixed to the 2FGL values. In the source model of GRB~080916C, we included only this burst by modeling its spectrum with a single power-law function. Galactic and extragalactic diffuse emissions were modeled with the recommended template files ({\tt gal\_2yearp7v6\_v0.fits} and {\tt iso\_p7v6source.txt}), where the normalizations of both components were set free and the standard power-law scaling was applied to the Galactic one. As a result, we obtained a photon index of $\alpha=2.08\pm0.06$ and a curvature parameter $\beta=0.10\pm0.03$ for 2FGL~J1344-1723 and $\Gamma=2.18\pm0.08$ for GRB~080916C. To construct the LAT spectrum, we divided the whole energy range into bins defined by energies of 0.1--0.3, 0.3--1.0, 1.0--3.0, 3--10, 10--30, and 30--100~GeV and performed a maximum-likelihood analysis using {\tt gtlike} in each energy bin. We assumed single power-law shapes for both sources allowing the normalization to vary. Here the power law index was fixed to the values obtained by {\tt gtlike} from the whole energy-range (0.1--300 GeV) data set for GRB~080916C, while we utilize the $\alpha$ value of the log-parabola spectrum for 2FGL~J1344-1723. For the surrounding sources and the two (Galactic and isotropic) diffuse components, the spectral parameters were again fixed to the values obtained by {\tt gtlike} for the whole energy range and only the normalizations were left free. The estimated systematic uncertainty on the $\gamma$-ray flux is 10\% at 100~MeV, decreasing to 5\% at 560~MeV, and increasing to 10\% at 10~GeV and above \citep{ack12}.

\begin{figure}[t]
\plotone{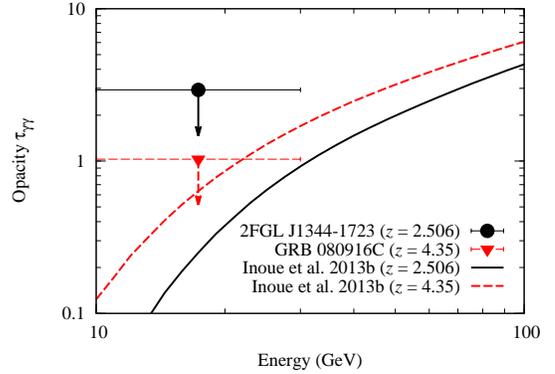} 
\caption{Derived upper limits for the optical depth at $z = 2.506$ (2FGL J1344-1723, solid arrow) and, $z = 4.35$ (GRB 080916C, dashed arrow). For the comparison, we also plot an expected optical depth from an EBL model by curves \citep{ino13}. \label{fig:tau} }
\end{figure}

To constrain the gamma-ray opacity, an intrinsic spectrum is required. There are various models to explain blazar and GRB spectra. We follow general arguments based on the shock acceleration scenario. First-order Fermi acceleration is the broadly accepted model, and the resulting particle spectrum is expected to show a power-law shape of $dN/dE \propto E^{-p}$ with a slope $p\sim2$. The slope $p$ can be steeper than 2 due to cooling effects. The expected photon index from the inverse Compton scattering radiation by electrons will be $\Gamma\ge(1+p)/2$, while that from hadronic processes will be $\Gamma=p$, where the photon spectrum is given by $dN_\gamma/dE\propto E^{-\Gamma}$. The resulting photon index of the intrinsic spectrum is $\Gamma\ge1.5$ \citep[e.g.][]{aha06}.  If we assume a softer intrinsic spectrum, we can obtain a tighter constraint on Pop~III star formation rate density. To be conservative, we take $\Gamma=1.5$ \citep[e.g.][]{aha06} from the energy bin of 3.0--10.0 GeV, which is the highest energy bin among the data points not affected by the gamma-ray attenuation. The gamma-ray opacity at energy $E$ and redshift $z$ is estimated as $\tau_{\gamma\gamma}(E,z)=\ln[F_{\rm int}(E)/F_{\rm obs}(E)]$, where $F_{\rm int}$ is the intrinsic flux and $F_{\rm obs}$ is the observed flux. We take this opacity as an upper limit. Figure \ref{fig:tau} shows the derived opacity limit for 2FGL~J1344-1723 and GRB~080916C. The derived upper limit at $E=17.3$~GeV is $\tau_{\gamma\gamma}\le2.93$ and $\le1.03$ at $z=2.506$ and $z=4.35$, respectively.  We also show the opacity from the model of \citet{ino13} for reference. We note that hadron-induced gamma rays may appear above several hundred GeV \citep{ess10}.

In summary, constraints on Pop~III stars obtained by current available observations are $\tau_e\le0.092$ \citep{ade13}, $x_{\rm HI}\ge0.033$ at $z=6.2$ \citep{mes07}, $x_{\rm HI}\ge0.1$ at $z=7.08$ \citep{mor11}, $\tau_{\gamma\gamma}\le2.93$ at $E=17.3$~GeV and $z=2.506$, and $\tau_{\gamma\gamma}\le1.03$ at $E=17.3$~GeV and $z=4.35$. Since probability functions of these limits are uncertain, we set the upper limits by taking the maximum allowed star formation rate density at each redshift within each allowed parameter spaces. Therefore, first, we estimate $\tau_e$, $x_{\rm HI}$ and $\tau_{\gamma\gamma}$ in all the combinations of parameters. Second, we choose parameter sets that do not violate the given observational constraints. Lastly, we take the maximum of the star formation rate density at each redshift from the allowed parameter sets and set those as the upper limit on the Pop~III star formation history.

\section{Results}

\begin{figure}[t]
\plotone{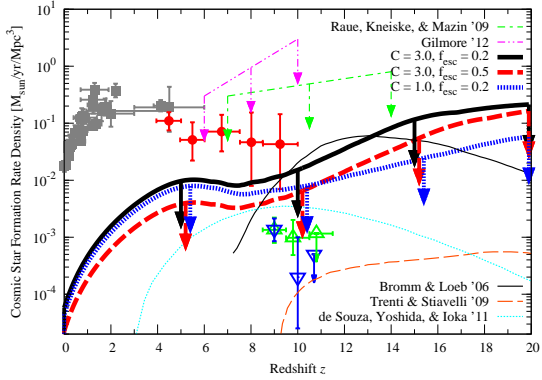} 
\caption{Upper limits on the Pop~III star formation history. Thick solid, dashed, and dotted curve with arrows represents the upper limit for $(C, f_{\rm esc})=(3.0, 0.2)$, $(3.0, 0.5)$, and $(1.0, 0.2)$, respectively. For comparison, we show the previously studied upper limits by  \citet{rau09} and \citet{gil11} with thin dot-dashed line and double-dot-dashed line with arrows. We plot theoretical models for the Pop~III star formation by \citet{bro06}, thin solid curve; \citet{tre09} (Pop~III.2), dashed curve; and \citet{des11} (Pop~III.2), dot-dashed curve. For \citet{des11}, we adopt their optimistic model. We also plot the observational data compiled by \citet[][filled square]{hop04}, that deduced from high redshift galaxies \citep[][up-triangle, and down-triangle, respectively]{ell13,oes13}, and that inferred from GRBs \citep[][circle]{kis13}. These data points represent the summation of all stellar populations. For the high redshift galaxies, limiting luminosities are adopted by each author.
 \label{fig:csfh} }
\end{figure}

Figure \ref{fig:csfh} shows the upper limits on the Pop~III star formation history. Thick solid, dashed, and dotted curves with arrows show the upper limit on the Pop~III star formation rate density for $(C, f_{\rm esc})=(3.0, 0.2)$, (3.0, 0.5), and (1.0, 0.2), respectively. The upper limit above $z=7$ can be approximated as 
\begin{equation}
\dot\rho_{*}(z)<0.01\left[\frac{(1+z)}{(1+7.0)}\right]^{3.4}\left(\frac{f_{\rm esc}}{0.2}\right)^{-1}\left(\frac{C}{3.0}\right) \ {\rm M}_{\odot}\ {\rm yr}^{-1}\ {\rm Mpc}^{-3}.
\end{equation}
For comparison, we show the upper-limits from previous studies \citep{rau09,gil11}, which do not take into account the reionization constraints.  The upper limit becomes 10 times tighter by including the reionization data. The upper limit does not change significantly when we exclude the constraints from the {\it Fermi}-LAT data. This is because the most distant gamma-ray objects with $>10$ GeV photons are still not beyond $z\sim6$ where the reionization occurred.  

We also plot the cosmic star formation rate density data compiled by \citet{hop04}, deduced from distant galaxies \citep{ell13,oes13}, and inferred from GRBs \citep{kis13}. We note that these data represent all stellar populations. Each high redshift galaxy data point is obtained by integrating its luminosity function down to a limiting luminosity. We note that  the overall UV luminosity function at these redshifts is still uncertain due to poor statistics \citep{ell13,oes13}.

We compare our limits with theoretically studied Pop~III star formation rates in the literature, specifically in \citet[][Pop~III.1 + III.2]{bro06}, \citet[][Pop~III.2]{tre09}, and \citet[][Pop~III.2]{des11}. For \citet{des11}, we take their optimistic model where they assume a very high star formation efficiency and low chemical enrichment. The Pop~III star formation model by \citet{bro06} violates the current observational constraints. 

If remnants of Pop~III stars became seeds of supermassive black holes (SMBHs), which would manifest themselves as quasars observed at $z\sim6$ \citep[e.g.][]{mor11}, the presence of such quasars may tighten our limits. In this Letter, we do not take into account this constraint because of our poor knowledge of accretion and merger history of such seed BHs, and their precursors. It may be difficult to create SMBHs from Pop~III stars because radiation feedback terminates Pop~III stars growing above a few tens solar masses \citep[e.g.][]{hos11}. Promising candidates for seeds of SMBHs can be formed by a direct collapse of supermassive stars \citep{hos13}. Since their effective temperature is less than $10^4$ K because of large radius of such supermassive stars, they will not significantly contribute to the reionization.

\section{Conclusions}
\label{sec:con}
In this Letter, we show that considering simultaneously the current measurements of gamma-ray opacity, the electron Thomson scattering optical depth, and the neutral fraction set an upper limit of $\dot\rho_{*}(z)<0.01[(1+z)/{(1+7.0)}]^{3.4}({f_{\rm esc}}/{0.2})^{-1}({C}/{3.0})\ {\rm M}_{\odot}\ {\rm yr}^{-1}\ {\rm Mpc}^{-3}$ on the Pop~III star formation rate density above $z=7$. By including the reionization constraints, the upper limit becomes $\sim10$ times tighter compared to previous works \citep{rau09,gil11}.

Current gamma-ray data do not strongly constrain the Pop~III star formation history because the most distant object is still at $z=4.35$ where the reionization process is completed. To explore the reionization epoch via the gamma-ray technique, sources beyond redshift 6 and rich with $>10$ GeV photons would be needed \citep{ino13}. {\it Fermi} may eventually detect blazars at $z>6$ \citep{ino10_highz}, and the Cherenkov Telescope Array may possibly do the same at $\sim30$ GeV for GRBs \citep{sin13_cta}. \citet{tan13} have reported the detection of $>$100 GeV photons from a blazar at $z=1.1$ and  \citet{tak13} have reported a candidate {\it Fermi} gamma-ray blazar at $z\sim3-4$. Future gamma-ray observations of distant sources at $\gtrsim10$ GeV will put tighter constraints on Pop~III star formation history by combining reionization measurements.

%% Acknowledgement
\acknowledgements
We acknowledge M. Ajello, J. Chiang, L. Latronico, P. Marshall, J. Perkins, J. Scargle, and D. Thompson for useful comments. The {\it Fermi}-LAT Collaboration acknowledges support 
from a number of agencies and institutes for both development and the operation of 
the LAT as well as scientific data analysis. These include NASA and DOE in the 
United States, CEA/Irfu and IN2P3/CNRS in France, ASI and INFN in Italy, MEXT, KEK, 
and JAXA in Japan, and the K.~A.~Wallenberg Foundation, the Swedish Research Council 
and the National Space Board in Sweden. Additional support from INAF in Italy and CNES 
in France for science analysis during the operations phase is also gratefully acknowledged.


\begin{thebibliography}{55}
\expandafter\ifx\csname natexlab\endcsname\relax\def\natexlab#1{#1}\fi

\bibitem[{{Abel} {et~al.}(2002){Abel}, {Bryan}, \& {Norman}}]{abe02}
{Abel}, T., {Bryan}, G.~L., \& {Norman}, M.~L. 2002, Science, 295, 93

\bibitem[{{Abramowski} {et~al.}(2013)}]{abr13_ebl}
{Abramowski}, A. {et~al.} 2013, \aap, 550, A4

\bibitem[{{Ackermann} {et~al.}(2012{\natexlab{a}})}]{ack12}
{Ackermann}, M. {et~al.} 2012{\natexlab{a}}, \apjs, 203, 4

\bibitem[{{Ackermann} {et~al.}(2012{\natexlab{b}})}]{ack12_ebl}
---. 2012{\natexlab{b}}, Science, 338, 1190

\bibitem[{{Ackermann} {et~al.}(2013)}]{ack13_grb}
---. 2013, arXiv:1303.2908

\bibitem[{{Ade} {et~al.}(2013)}]{ade13}
{Ade}, P.~A.~R. {et~al.} 2013, arXiv:1303.5076

\bibitem[{{Aharonian} {et~al.}(2006)}]{aha06}
{Aharonian}, F. {et~al.} 2006, \nat, 440, 1018

\bibitem[{{Bolton} \& {Haehnelt}(2007)}]{bol07}
{Bolton}, J.~S. \& {Haehnelt}, M.~G. 2007, \mnras, 382, 325

\bibitem[{{Bromm} \& {Loeb}(2006)}]{bro06}
{Bromm}, V. \& {Loeb}, A. 2006, \apj, 642, 382

\bibitem[{{Bromm} {et~al.}(2009){Bromm}, {Yoshida}, {Hernquist}, \&
  {McKee}}]{bro09}
{Bromm}, V., {Yoshida}, N., {Hernquist}, L., \& {McKee}, C.~F. 2009, \nat, 459,
  49

\bibitem[{{de Souza} {et~al.}(2011){de Souza}, {Yoshida}, \& {Ioka}}]{des11}
{de Souza}, R.~S., {Yoshida}, N., \& {Ioka}, K. 2011, \aap, 533, A32

\bibitem[{{Dom{\'{\i}}nguez} {et~al.}(2013){Dom{\'{\i}}nguez}, {Finke},
  {Prada}, {Primack}, {Kitaura}, {Siana}, \& {Paneque}}]{dom13}
{Dom{\'{\i}}nguez}, A., {Finke}, J.~D., {Prada}, F., {Primack}, J.~R.,
  {Kitaura}, F.~S., {Siana}, B., \& {Paneque}, D. 2013, \apj, 770, 77

\bibitem[{{Dom{\'{\i}}nguez} {et~al.}(2011)}]{dom11}
{Dom{\'{\i}}nguez}, A. {et~al.} 2011, \mnras, 410, 2556

\bibitem[{{Dwek} \& {Krennrich}(2013)}]{dwe13}
{Dwek}, E. \& {Krennrich}, F. 2013, Astroparticle Physics, 43, 112

\bibitem[{{Ellis} {et~al.}(2013)}]{ell13}
{Ellis}, R.~S. {et~al.} 2013, \apjl, 763, L7

\bibitem[{{Essey} \& {Kusenko}(2010)}]{ess10}
{Essey}, W. \& {Kusenko}, A. 2010, Astroparticle Physics, 33, 81

\bibitem[{{Fernandez} \& {Komatsu}(2006)}]{fer06}
{Fernandez}, E.~R. \& {Komatsu}, E. 2006, \apj, 646, 703

\bibitem[{{Gilmore}(2012)}]{gil11}
{Gilmore}, R.~C. 2012, \mnras, 420, 800

\bibitem[{{Gong} \& {Cooray}(2013)}]{gon13}
{Gong}, Y. \& {Cooray}, A. 2013, \apjl, 772, L12

\bibitem[{{Gould} \& {Schr{\'e}der}(1966)}]{gou66}
{Gould}, R.~J. \& {Schr{\'e}der}, G. 1966, Physical Review Letters, 16, 252

\bibitem[{{Gunn} \& {Peterson}(1965)}]{gun65}
{Gunn}, J.~E. \& {Peterson}, B.~A. 1965, \apj, 142, 1633

\bibitem[{{Hopkins}(2004)}]{hop04}
{Hopkins}, A.~M. 2004, \apj, 615, 209

\bibitem[{{Hosokawa} {et~al.}(2011){Hosokawa}, {Omukai}, {Yoshida}, \&
  {Yorke}}]{hos11}
{Hosokawa}, T., {Omukai}, K., {Yoshida}, N., \& {Yorke}, H.~W. 2011, Science,
  334, 1250

\bibitem[{{Hosokawa} {et~al.}(2013){Hosokawa}, {Yorke}, {Inayoshi}, {Omukai},
  \& {Yoshida}}]{hos13}
{Hosokawa}, T., {Yorke}, H.~W., {Inayoshi}, K., {Omukai}, K., \& {Yoshida}, N.
  2013, \apj, 778, 178

\bibitem[{{Hosokawa} {et~al.}(2012){Hosokawa}, {Yoshida}, {Omukai}, \&
  {Yorke}}]{hos12}
{Hosokawa}, T., {Yoshida}, N., {Omukai}, K., \& {Yorke}, H.~W. 2012, \apjl,
  760, L37

\bibitem[{{Inoue} {et~al.}(2010){Inoue}, {Salvaterra}, {Choudhury}, {Ferrara},
  {Ciardi}, \& {Schneider}}]{sin10}
{Inoue}, S., {Salvaterra}, R., {Choudhury}, T.~R., {Ferrara}, A., {Ciardi}, B.,
  \& {Schneider}, R. 2010, \mnras, 404, 1938

\bibitem[{{Inoue} {et~al.}(2013{\natexlab{a}})}]{sin13_cta}
{Inoue}, S. {et~al.} 2013{\natexlab{a}}, Astroparticle Physics, 43, 252

\bibitem[{{Inoue} {et~al.}(2013{\natexlab{b}}){Inoue}, {Inoue}, {Kobayashi},
  {Makiya}, {Niino}, \& {Totani}}]{ino13}
{Inoue}, Y., {Inoue}, S., {Kobayashi}, M.~A.~R., {Makiya}, R., {Niino}, Y., \&
  {Totani}, T. 2013{\natexlab{b}}, \apj, 768, 197

\bibitem[{{Inoue} {et~al.}(2011){Inoue}, {Inoue}, {Kobayashi}, {Totani},
  {Kataoka}, \& {Sato}}]{ino10_highz}
{Inoue}, Y., {Inoue}, S., {Kobayashi}, M.~A.~R., {Totani}, T., {Kataoka}, J.,
  \& {Sato}, R. 2011, \mnras, 411, 464

\bibitem[{{Jelley}(1966)}]{jel66}
{Jelley}, J.~V. 1966, Physical Review Letters, 16, 479

\bibitem[{{Johnson} {et~al.}(2008){Johnson}, {Greif}, \& {Bromm}}]{joh08}
{Johnson}, J.~L., {Greif}, T.~H., \& {Bromm}, V. 2008, \mnras, 388, 26

\bibitem[{{Kistler} {et~al.}(2013){Kistler}, {Yuksel}, \& {Hopkins}}]{kis13}
{Kistler}, M.~D., {Yuksel}, H., \& {Hopkins}, A.~M. 2013, arXiv:1305.1630

\bibitem[{{Madau} \& {Silk}(2005)}]{mad05}
{Madau}, P. \& {Silk}, J. 2005, \mnras, 359, L37

\bibitem[{{Matsumoto} {et~al.}(2005)}]{mat05}
{Matsumoto}, T. {et~al.} 2005, \apj, 626, 31

\bibitem[{{Mesinger}(2010)}]{mes10}
{Mesinger}, A. 2010, \mnras, 407, 1328

\bibitem[{{Mesinger} \& {Haiman}(2007)}]{mes07}
{Mesinger}, A. \& {Haiman}, Z. 2007, \apj, 660, 923

\bibitem[{{Mortlock} {et~al.}(2011)}]{mor11}
{Mortlock}, D.~J. {et~al.} 2011, \nat, 474, 616

\bibitem[{{Nestor} {et~al.}(2013){Nestor}, {Shapley}, {Kornei}, {Steidel}, \&
  {Siana}}]{nes13}
{Nestor}, D.~B., {Shapley}, A.~E., {Kornei}, K.~A., {Steidel}, C.~C., \&
  {Siana}, B. 2013, \apj, 765, 47

\bibitem[{{Nolan} {et~al.}(2012)}]{nol12_catalog}
{Nolan}, P.~L. {et~al.} 2012, \apjs, 199, 31

\bibitem[{{Oesch} {et~al.}(2013)}]{oes13}
{Oesch}, P.~A. {et~al.} 2013, arXiv:1301.6162

\bibitem[{{Oh}(2001)}]{oh01}
{Oh}, S.~P. 2001, \apj, 553, 25

\bibitem[{{Ono} {et~al.}(2010){Ono}, {Ouchi}, {Shimasaku}, {Dunlop}, {Farrah},
  {McLure}, \& {Okamura}}]{ono10}
{Ono}, Y., {Ouchi}, M., {Shimasaku}, K., {Dunlop}, J., {Farrah}, D., {McLure},
  R., \& {Okamura}, S. 2010, \apj, 724, 1524

\bibitem[{{Pritchard} {et~al.}(2010){Pritchard}, {Loeb}, \& {Wyithe}}]{pri10}
{Pritchard}, J.~R., {Loeb}, A., \& {Wyithe}, J.~S.~B. 2010, \mnras, 408, 57

\bibitem[{{Raue} {et~al.}(2009){Raue}, {Kneiske}, \& {Mazin}}]{rau09}
{Raue}, M., {Kneiske}, T., \& {Mazin}, D. 2009, \aap, 498, 25

\bibitem[{{Salpeter}(1955)}]{sal55}
{Salpeter}, E.~E. 1955, \apj, 121, 161

\bibitem[{{Schaerer}(2003)}]{sch03}
{Schaerer}, D. 2003, \aap, 397, 527

\bibitem[{{Shull} \& {Venkatesan}(2008)}]{shu08}
{Shull}, J.~M. \& {Venkatesan}, A. 2008, \apj, 685, 1

\bibitem[{{Takahashi} {et~al.}(2013){Takahashi}, {Kataoka}, {Niinuma}, {Honma},
  {Inoue}, {Totani}, {Inoue}, {Nakamori}, \& {Maeda}}]{tak13}
{Takahashi}, Y., {Kataoka}, J., {Niinuma}, K., {Honma}, M., {Inoue}, Y.,
  {Totani}, T., {Inoue}, S., {Nakamori}, T., \& {Maeda}, K. 2013,
  arXiv:1306.3552

\bibitem[{{Tanaka} {et~al.}(2013)}]{tan13}
{Tanaka}, Y.~T. {et~al.} 2013, arXiv:1308.0595

\bibitem[{{Tornatore} {et~al.}(2007){Tornatore}, {Ferrara}, \&
  {Schneider}}]{tor07}
{Tornatore}, L., {Ferrara}, A., \& {Schneider}, R. 2007, \mnras, 382, 945

\bibitem[{{Totani} {et~al.}(2006){Totani}, {Kawai}, {Kosugi}, {Aoki}, {Yamada},
  {Iye}, {Ohta}, \& {Hattori}}]{tot06}
{Totani}, T., {Kawai}, N., {Kosugi}, G., {Aoki}, K., {Yamada}, T., {Iye}, M.,
  {Ohta}, K., \& {Hattori}, T. 2006, \pasj, 58, 485

\bibitem[{{Trenti} \& {Stiavelli}(2009)}]{tre09}
{Trenti}, M. \& {Stiavelli}, M. 2009, \apj, 694, 879

\bibitem[{{Tsumura} {et~al.}(2013){Tsumura}, {Matsumoto}, {Matsuura}, {Sakon},
  \& {Wada}}]{tsu13}
{Tsumura}, K., {Matsumoto}, T., {Matsuura}, S., {Sakon}, I., \& {Wada}, T.
  2013, arXiv:1307.6740

\bibitem[{{Yajima} {et~al.}(2011){Yajima}, {Choi}, \& {Nagamine}}]{yaj11}
{Yajima}, H., {Choi}, J.-H., \& {Nagamine}, K. 2011, \mnras, 412, 411

\bibitem[{{Yoshii} \& {Peterson}(1994)}]{yos94}
{Yoshii}, Y. \& {Peterson}, B.~A. 1994, \apj, 436, 551

\end{thebibliography}
\end{document}